# Structural Engineering of Colloidal Quantum Dots: Towards Realization of Highly Efficient, Aerobic-Stable, and Droop-Free QLEDs

*Sadra Sadeghi, Saeedeh Mokarian Zanjani, Sergey Dayneko, Christian J. Imperiale, Francesco Tintori, Stéphane Kéna-Cohen, Afshin Shahalizad, and Majid Pahlevani\**

S. Sadeghi, S. Mokarian Zanjani, F. Tintori, and M. Pahlevani
Department of Electrical and Computer Engineering
Queen's University
Kingston, Ontario K7L 3N6, Canada
E-mail: majid.pahlevani@queensu.ca

A. Shahalizad and S. Dayneko
Printed Electronics Division
Genoptic LED Inc.
Calgary, Alberta T2C 5C3, Canada

C.J. Imperiale and S. Kéna-Cohen
Department of Engineering Physics
École Polytechnique de Montréal
PO Box 6079, succ. Centre-Ville, Montreal QC H3C 3A7, Canada

S. Sadeghi and S. Mokarian Zanjani contributed equally to this work.

# Abstract

Quantum dot light-emitting diodes (QLEDs) are promising building blocks for prospective lighting and display applications. Despite the significant advancements achieved towards increasing the efficiency and brightness levels of QLEDs, the unavoidable demand for an inert atmosphere during the fabrication process restrains their potential for large-scale manufacturing. Here, we demonstrate ZnCdSe/ZnSe/ZnSeS/ZnS core/multi-shell QDs with ultra-thick shell (14 monolayers), significantly suppressing the diffusion of moisture and oxidative species to realize efficient all solution-processed QLED devices fully fabricated under ambient air conditions. The lattice-mismatch-engineered shell growth in the synthesized QDs leads to the strong confinement of charge carriers inside the core and near-unity photoluminescence quantum yield (PLQY) in the amber-to-red wavelength region (595-625 nm). The best amber-emitting QLED devices exhibit an external quantum efficiency (EQE) of 14.8%, a current efficiency (CE) of 41.1 cd/A, and a luminance of $5.6 \times 10^5$ cd/m$^2$ at 12 V. To the best of our knowledge, these are the highest efficiency levels achieved in air-fabricated QLEDs. Importantly, the demonstrated devices show significant suppression of non-radiative Auger recombination and optimal charge balance, leading to negligible EQE roll-off (i.e., droop-free behavior) at the maximum luminance level. Furthermore, employing a refractive-index-matched optical compound to enhance light out-coupling increases the EQE to 26.6%, which is comparable to the best devices fabricated in an inert atmosphere. Collectively, these advances will greatly facilitate the large-scale industrial fabrication of air-stable QLEDs.

# 1. Introduction

Semiconductor quantum dots (QDs) offer significant potential for reshaping the future of optoelectronic devices. These materials feature unique size-dependent emission wavelength due to the 'quantum confinement' effect, high photoluminescence quantum yield (PLQY), narrow luminescence bandwidth, excellent color purity, and high photochemical stability.[1, 2] Owing to their exceptional optoelectronic properties, QDs are highlighted in a vast spectrum of applications ranging from energy harvesting,[3, 4] bioimaging and biodiagnostic,[5] communication,[6] and light-emitting diodes for illumination and display.[7] Since the first demonstration of QD-based light-emitting diodes (QLEDs),[8] Optoelectronic properties of colloidal QDs and device architecture are continuously engineered to realize more efficient devices with purer colors.[9-13] As a result, the efficiency levels of QLEDs are now comparable with the state-of-the-art organic LEDs (OLEDs).[14, 15] Lately, in parallel with the optimization of the charge transport layers, efforts have mostly shifted towards the engineering of novel QD heterostructures, which could also resolve the issues associated with short device stability and severe efficiency roll-off in QLEDs.[16] These optimizations are primarily focused on suppressing surface defects as the main source of charge carrier trapping and non-radiative Auger recombination.[17] To effectively passivate the surface defects, engineering of the surface ligands,[18, 19] designing new core/shell structures for better charge confinement,[20, 21] and engineering of the core/shell interface by utilizing the alloyed composition[22, 23] have been implemented so far.

To obtain high QLED efficiency and stability, QDs need to possess high PLQYs in thin films and maintain charge balance within devices.[24] The simplest conventional design of red-emitting core/shell QDs is typically composed of a CdSe core covered with large-bandgap inorganic compounds such as ZnS for better confinement of excitons inside the core (Figure 1a). However, the large lattice mismatch between the CdSe core and ZnS shell (~12%) induces interfacial surface defects,[25] leading to charge trapping and low device efficiency. At the same time, due to the proximity of the conduction band in ZnS and ZnO (typically in the range of 0.2 eV),[26] electron transfer is highly favored in comparison with the hole transfer, which leads to significant interfacial exciton quenching. As an alternative strategy, an interlayer of CdS[27] or ZnSe[7] was introduced, which lowered the lattice mismatch and graded the confinement potential (Figure 1b,c). However, in core/shell QDs with reduced lattice mismatch yet with abrupt interfaces, the formation of interface defects by increasing the shell thickness is inevitable.[17] The QDs with alloyed compositions offer (i) finely-tuned lattice mismatch, which mitigates the formation of interfacial defects between the core and shell materials[28] and (ii) a gradual change of confinement potential, which improves exciton confinement and suppresses non-radiative Auger recombination (Figure 1d).[29] Several types of QDs with alloyed compositions have already been investigated in QLEDs, showing superior efficiency and brightness levels in comparison with conventional core/multi-shell QDs.[14, 15, 23, 29-31] Record external quantum efficiencies (EQEs) of 30.9%, 28.7%, and 21.9% have been achieved for solution-processed QLEDs based on red-emitting ZnCdSe/ZnSe/ZnS,[31] green-emitting CdSe/CdZnSe/ZnS, and blue-emitting CdZnSe/ZnS QDs,[32] respectively. Previously, Shen and co-workers introduced the novel ZnCdSe/ZnSe/ZnSeS/ZnS heterostructure with a high peak EQE of 23.9% in the green spectral region.[33] In this structure, the diffusion of Zn into the ZnCdSe core during the ZnSe shell growth blue-shifted the PL peak towards the high-energy portion of the visible spectrum. However, achieving a PL peak in the amber-to-red spectral region requires a precise core and shell tailoring strategy, which has not been investigated so far for this type of QDs.

The graded composition in 'giant' ZnCdSe/ZnSe/ZnSeS/ZnS QDs could be beneficial to the aerobic stability of the QLED by facilitating charge injection and preventing the diffusion of moisture and oxidative species to the core.[34, 35] Up until now, the vast majority of the efficient QLED devices are fabricated under inert atmosphere (e.g. nitrogen-filled glove-box) due to the fast degradation of the organic layers, ZnO ETL, and QDs, which deteriorates the efficiency and stability.[36] To the best of our knowledge, there is currently no report in the literature on efficient fully in-air fabricated QLEDs. Recently, Ren and co-workers showed the rapid decay of the luminance for the fully in-air fabricated QLEDs. These devices dropped under 20% of the initial luminance after 60 minutes in comparison with the devices manufactured in the glove-box, which preserved more than 90% of their initial luminance during the same time.[37] In another study by Sun and co-workers, deposition of ZnMgO in air showed a much lower EQE (4.63%) in comparison with the QLED devices fully fabricated in glove-box (21.87%), due to the significant adsorption of oxygen by ZnMgO layer, which was found to reduce the radiative recombination efficiency and lowered electron transport.[38] The apparent need for inert atmosphere to realize the highly-efficient solution-processed QLED devices limits their application in large-scale manufacturing processes. Therefore, demonstrating highly-efficient QLEDs fully fabricated under ambient air conditions is essential for their widespread use in future lighting applications. The graded composition of the core/multi-shell QDs also effectively suppresses the efficiency roll-off (i.e., efficiency droop) in QLED devices.[29] The efficiency droop is the progressive deterioration of the QLED performance with increasing the current or lumiannce. Importantly, the EQE droop typically observed in QLEDs limits their maximum achievable luminance levels and the device longevity, due to the rapid transfer of the exciton recombination energy to the extra charge carriers instead of being released as a photon.[39] Hence, controlling the droop effect in QLEDs could beneficially increase their operational stability by decreasing the generated heat in QLEDs, making them suitable particularly for high-brightness (>5000 cd.m$^{-2}$) outdoor lighting applications.

Herein, we synthesized a novel type of near-unity-emitting (PLQY of up to 98%) ZnCdSe/ZnSe/ZnSeS/ZnS QDs with narrow PL linewidths (<30 nm), covering the amber-to-red wavelength region (595-625 nm). The synthesized core/multi-shell QDs showed strong confinement of charge carriers by the lattice mismatch-suppressed graded-shell configuration. Notably, a robust aerobic stability during device fabrication process was obtained owing to the ultra-thick shell formation up to 14 monolayers (MLs). The improved aerobic stability of the synthesized 'giant' QDs (average size of 15.1 nm) led to a maximum EQE of 14.8%, a current efficiency (CE) of 41.1 cd.A$^{-1}$, a power efficiency (PE) of 31.8 lm.W$^{-1}$, and a luminance of 560,528 cd.m$^{-2}$ at 12 V. To the best of our knowledge, these are the highest efficiency levels achieved in fully air-fabricated devices. The engineered graded structure enabled a smooth charge injection barrier in the QLED devices together with a significant suppression of Auger recombination, leading to nearly-droop-free behavior (<7%) at the maximum luminance level. Finally, a refractive-index-matched immersion oil was used to enhance the light out-coupling efficiency of the QLED devices, which boosted the EQE level up to 26.6%, with a luminance of more than 1,000,000 cd.m$^{-2}$, PE of 53.5 lm.W$^{-1}$, and CE of 73.1 cd.A$^{-1}$.

## 2. Results and Discussion

To synthesize the giant ZnCdSe/ZnSe/ZnSeS/ZnS QDs with alloyed composition, we modified the procedure based on a previous study, which showed near-unity emitting ZnCdSe/ZnSe QDs in the red spectral region (PL peak at 631 nm).[40] In this procedure, upon 'fast' injection of the Se precursor into the reaction flask containing $Cd^{+2}$ and $Zn^{+2}$ cations at high temperature (300 °C), the nucleation and growth of CdSe cores lead to a wide distribution of the formed cores and broad full-width-at-half maximum (FWHM). At the same time, a thin layer of ZnSe shell starts to grow onto the formed CdSe cores, which further triggers the exchange reaction between $Zn^{+2}$ and $Cd^{+2}$ cations in the shell and core, respectively (Figure 1e). The continuous isotropic cation exchange reaction results in the formation of the alloyed ZnCdSe cores due to the reaction temperature being higher than the 'alloying point' at 270 °C, which significantly increases the diffusion coefficient.[41] Consequently, by continuous 'slow' injection of the Se precursor and the abundant Zn in the reaction, a relatively thick layer of ZnSe (9 ML) forms, which narrows the FWHM by hindering the 'Ostwald ripening' process (Figure 1e and 2b). Importantly, the injection of the Se precursor in the ZnSe shell formation should be sufficiently 'slow' to allow the epitaxial growth of the ZnSe shell onto the cores at high temperature, which significantly enhances the optical properties of the core/shell QDs.[42] To alleviate the large lattice mismatch between ZnSe and ZnS layers (approximately 5%),[33] an ultra-thin (1-2 ML) interlayer of alloyed ZnSeS was grown onto the ZnCdSe/ZnSe QDs. At the last step of the shelling process, approximately 4 ML of ZnS was grown onto the ZnCdSe/ZnSe/ZnSeS nanocrystals. Finally, for effective charge transfer to the active layer inside the device, we performed the ligand exchange procedure between the initial long-chain carboxylic acid (oleic acid) and the short-chain thiol ligand (1-octanethiol). Advantageously, it has been shown that the effective ligand exchange from oleate to thiol ligands results in enhanced electron mobility and balanced charge injection, leading to higher EQEs of QLED devices.[43, 44]

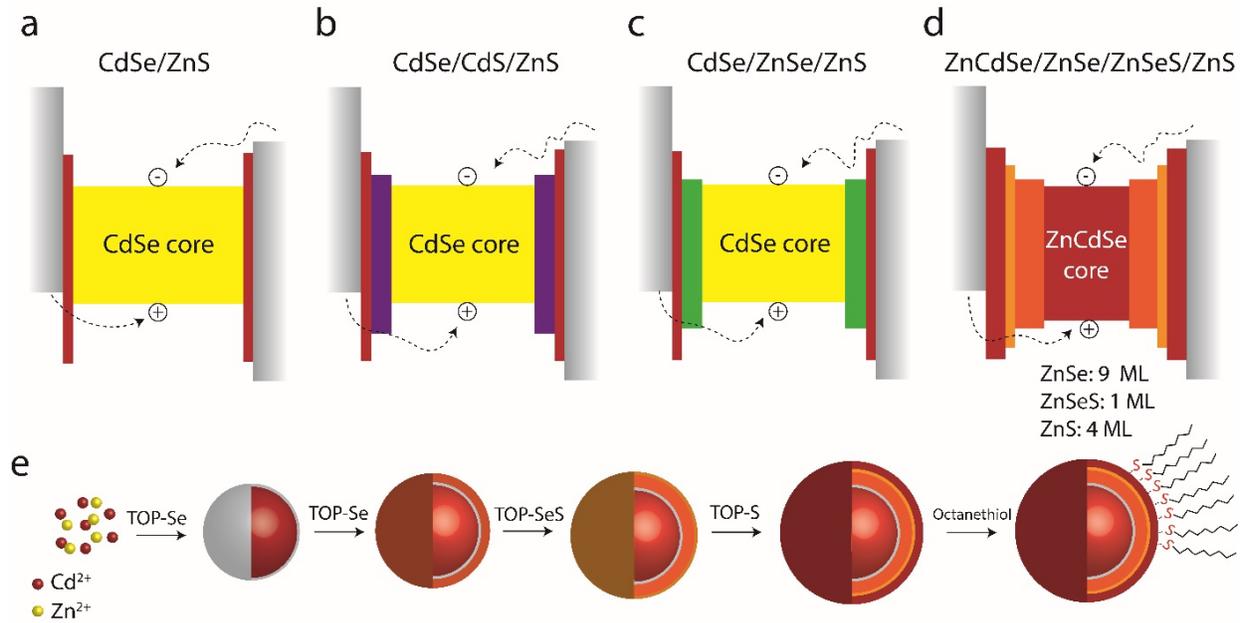

**Figure 1.** (a-c) Charge injection mechanism in QLED devices based on conventional quantum dot compositions reported in literature, and (d) the synthesized graded quantum dots in this work. (b) Schematic of the synthesis method to achieve highly efficient amber-to-red-emitting ZnCdSe/ZnSe/ZnSeS/ZnS QDs.

To obtain the PL peak of the ZnCdSe/ZnSe/ZnSeS/ZnS QDs in the amber-to-red wavelength region (595-625 nm), the evolution of the PL peak during the formation of the ZnCdSe core and each step of the shelling procedure needs to be carefully tuned. It is known that the higher reactivity of Se with Cd -rather than Zn- induces the formation of the CdSe cores QDs.[45] The continuous growth of the CdSe cores caused a red-shift of the PL peak to 641 nm (Figure 2a). At this stage, by further addition of Se, a thin layer of ZnSe formed and the Zn diffused inside the CdSe to form ZnCdSe core QDs (see Figure 3e). The simultaneous alloying process of the ZnCdSe and growth of the ZnSe shell led to a significant blue-shift of the PL peak (from 641 nm to 617 nm) and a notable narrowing of the FWHM from 34 nm to 21 nm (Figure 2a and 2b). The narrowing of the emission was previously attributed to the alloying process and uniform particle size distribution by the continuous rate of monomer production.[46] In our experiments, the growth of the thick ZnSe shell (9 ML) onto the alloyed ZnCdSe core QDs was also confirmed by the PLQY values, in which the sufficiently slow reaction time at high temperature (300 °C) during the addition of Se increased the PLQY from 42% to 75% for ZnCdSe cores and ZnCdSe/ZnSe QDs, respectively (Figure 2b). The PLQY enhancement continued by the formation of ZnSeS and ZnS shells and 1-octanethiol ligand treatment to the ultimate value of 98% for the final ZnCdSe/ZnSe/ZnSeS/ZnS-OT QDs (Figure 2b), which is due to the effective passivation of the core and strong confinement of the charge carriers. The PL peak of the full core/multi-shell structure (616 nm) did not show a significant wavelength shift from the initial ZnCdSe/ZnSe QDs (Figure 2a).

The PL peak of the synthesized ZnCdSe/ZnSe/ZnSeS/ZnS QDs could be conveniently fine-tuned to cover the whole amber-to-red region by adjusting the concentration of the initial Cd precursor. To this end, the Cd precursor concentration was varied from 0.25 to 0.4 mmol, red-shifting the PL peak from 595 nm to 624 nm, respectively (Figure 2c). The addition of the extra Cd (more than 0.4 mmol) significantly decreased both solution and film PLQY of the synthesized QDs (Figure S1, Supporting Information), ultimately leading to the low efficiency of the fabricated device. Similarly, decreasing the amount of Cd favored the excess Se in the reaction (Cd:Se, 0.25:0.5 molar ratio), at which the non-coordinated surface Se sites on the synthesized CdSe cores act as the deep hole trap states and reduce the PLQY[47] (Figure S1, Supporting Information). The emission red-shift observed upon increasing the Cd concentration was also reflected in the absorption spectra of the synthesized QDs (Figure 2d). Despite the high self-absorption in the synthesized QDs (Stokes-shift <10 meV), the solution PLQY values did not drop significantly (avg. 92% ± 6%) upon changing the Cd amount in the desired range. This was attributed to the strong confinement of electrons and holes inside the ZnCdSe core, leading to the efficient core emission instead of the emission from the other shell materials. Reducing the amount of Cd also improved the FWHM (from 29 to 21 nm) due to the formation of fewer nucleation sites and narrow size distribution (Figure 2e). In addition to high solution PLQY and narrow FWHM, the ZnCdSe/ZnSe/ZnSeS/ZnS QDs showed a PL red-shift of less than 4 nm in different batches when transferred from solution to film ensemble (Figure 2c). This small red-shift is likely due to the effective suppression of Förster resonance energy transfer (FRET) by the ultra-thick ZnSe/ZnSeS/ZnS shell. The significant suppression of FRET in the QDs films was also confirmed by the comparison of fluorescence lifetime decay of QD solution and film in time-resolved PL measurements (Figure 2f). The successful coating of the multi-shell on the synthesized ZnCdSe core QDs provides a smooth potential profile between the core and the shell, which effectively reduces the formation of exciton trapping centers in the interfaces, leading to a mono-exponential decay with the average lifetime of 16 ns for the ZnCdSe/ZnSe/ZnSeS/ZnS QD solution. The average lifetime only dropped by 25% (12 ns) in the film, suggesting the effective FRET suppression by multi-shell formation (Figure 2f).

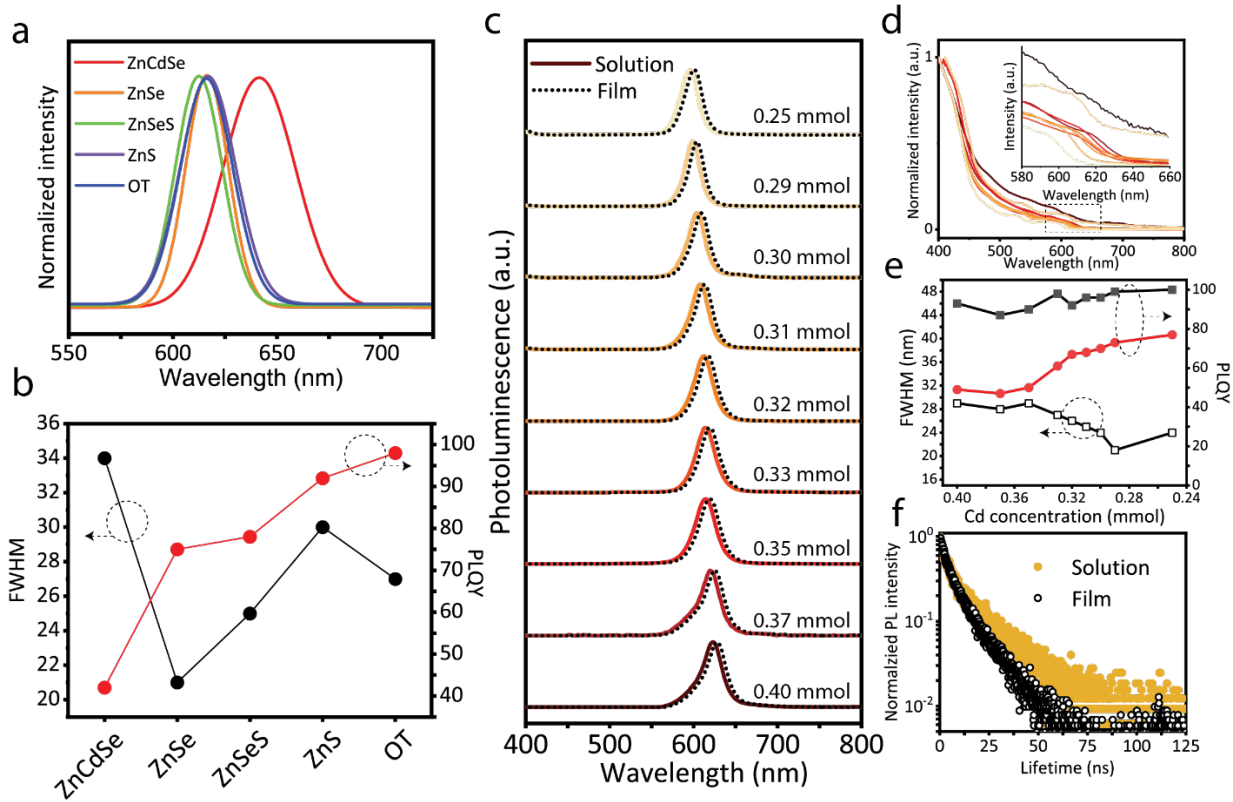

**Figure 2.** (a) Evolution of the PL spectra, (b) FWHM, and solution PLQY during ZnCdSe core formation, different steps of the shelling procedure, and ligand exchange. (c) Solution and film PL spectra, (d) absorbance spectra (inset: the zoomed area of the first excitonic peak), (e) solution (black squared line), film (red circled line) PLQY, and FWHM (black unfilled squared line) of ZnCdSe/ZnSe/ZnSeS/ZnS QDs by varying the cadmium content from 0.25 to 0.40 mmol in the reaction. (f) Solution and film PL decay curves of the synthesized amber-emitting QDs.

The morphology investigation of the synthesized QDs confirmed their nearly-spherical shape with the calculated average size of 15.1 ± 1.2 nm measured from transmission electron microscopy (TEM) image (Figure 3a). A size comparison between ZnCdSe core (Figure S2b, Supporting Information) and the core/multi-shell QD structures showed an increase in thickness upon the ZnSe/ZnSeS/ZnS shelling procedure to approximately 4.2 nm (equal to ~14 ML). Elemental distribution maps also demonstrated the successful coating of different layers (Figure 3b-f). Cd was mainly detected in the core, whereas Zn was distributed throughout the whole QD particles, confirming the presence of Zn in the ZnCdSe core and all shell layers. Due to the presence of Se in the core, ZnSe, and ZnSeS interlayers, distribution of Se was observed in the nanoparticles (NPs) with a slightly lower diameter in comparison with the Zn distribution (Figure 3d and 3e). The scattering of S at low content in the distribution map suggested the presence of S mainly in the thin ZnSeS shell and the ZnS outer shell (Figure 3f). Based on the distribution maps and the TEM images obtained from the ZnCdSe core and ZnCdSe/ZnSe/ZnSeS/ZnS QDs, the thickness of the ZnSe inner shell, ZnSeS interlayer, and ZnS outermost shell were measured as approximately 2.7 nm (9 ML), 0.4 nm (1 ML), and 1.1 nm (4 ML), respectively. From the high-resolution TEM (HRTEM) image, the surface of the QDs is partially shaded due to the presence of the organic ligand (Figure 3g). The HRTEM image exhibited continuous lattice fringes and no lattice mismatch throughout the particle, confirming the epitaxial growth of the shells onto the ZnCdSe core. The measured lattice interspacing was 0.3 nm, suggesting the formation of the cubic phase structure (Figure 3g-inset). The growth of the shells with cubic phase structure was further confirmed by x-ray diffraction (XRD) patterns (Figure 3h). In the XRD pattern of ZnCdSe core QDs, the diffraction peaks of the (111), (220), (311) appeared compatible with the diffraction peaks of the cubic CdSe and ZnSe (Figure 3h-lower panel), confirming the formation of the cubic alloyed structure. On the other hand, in the XRD pattern of ZnCdSe/ZnSe/ZnSeS/ZnS QDs, a low-intensity peak was distinguishable at 28.5°, suggesting the formation of the ZnS as the outermost layer. The slight displacement of the obtained peak positions in comparison with the standard XRD patterns of CdSe, ZnSe, and ZnS was due to the enlargement of the lattice structure by the formation of the alloyed composition.[48]

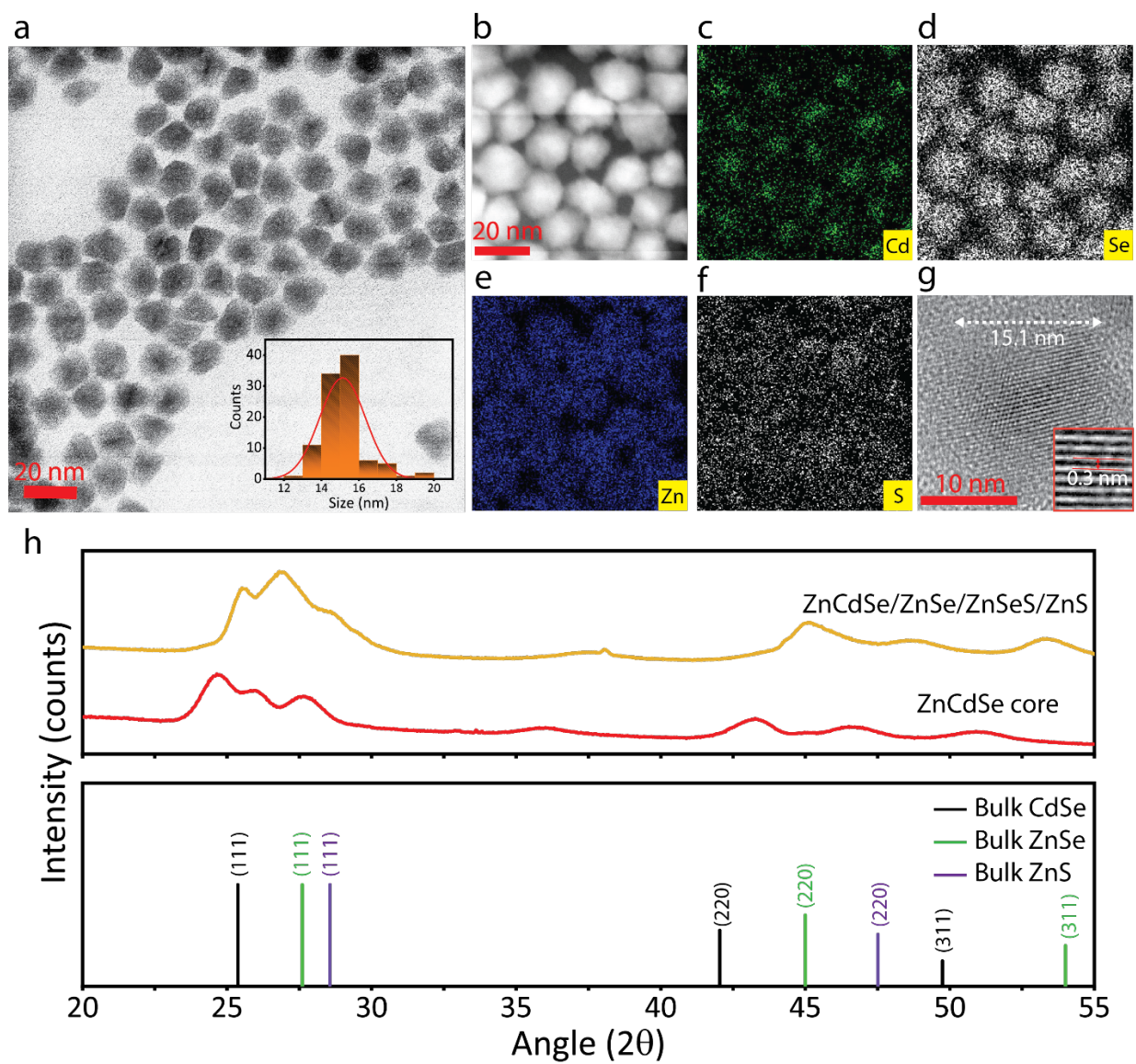

**Figure 3.** (a) TEM image of the synthesized ZnCdSe/ZnSe/ZnSeS/ZnS QDs (inset: size distribution of the QDs calculated from 100 particles). (b-f) Energy dispersive spectroscopy elemental maps of the Cd, Se, Zn, and S in ZnCdSe/ZnSe/ZnSeS/ZnS QDs. (g) HRTEM image of the synthesized core/multi-shell QDs (inset: demonstration of the lattice interspacing in the zoomed HRTEM image). (h) (Upper panel) the standard XRD pattern of the ZnCdSe core and ZnCdSe/ZnSe/ZnSeS/ZnS QDs. (Lower panel) position of the diffraction peaks for bulk CdSe (JCPDS card no. 18-0191), ZnSe (JCPDS card no. 01-088-2345), and ZnS (JCPDS card no. 05-0566).

QLED devices were then fabricated to further characterize the QDs and to assess their viability for lighting applications. The whole fabrication process was carried out in air and using solution processed layers (except for the top electrode). The fabricated QLED architecture is composed of the following layers: patterned indium tin oxide (ITO) anode (150 nm), poly(ethylene dioxythiophene): polystyrene sulfonate (PEDOT: PSS) as hole injection layer (HIL) (80 nm), a poly[(9,9-dioctylfluorenyl-2,7-diyl)-co-(4,4'-(N-(p-butylphenyl)) diphenylamine] (TFB) layer as HTL (40 nm), two MLs of QD film as the active layer (35 nm), LiZnMgO as ETL (110 nm), topped by aluminum (Al) (80 nm) and silver (Ag) (175 nm) as the cathode (Figure 4a,b and S4, Supporting Information). TFB is frequently used as HTL due to the higher hole mobility of ~$10^{-3}$ $cm^2.V^{-1}.s^{-1}$ in comparison with other transport materials.[49] For the ETL, ZnO was co-doped with 10% $Li^+$ and 10% $Mg^{2+}$, which improves the charge balance in the device by slightly upshifting the conduction band and slowing down the commonly observed excess electron current.[26] At the same time, the interstitial Zn sites are filled with $Li^+$ and $Mg^{2+}$ cations, which decrease the concentration of O-H at the oxide surface (i.e., exciton quenching sites), resulting in the higher device efficiency.[50] It is noted that the 'low-temperature' synthesis of the LiZnMgO (below 2 °C) resulted in the nanoparticles with a small size (less than 3 nm based on the absorbance peak at 294 nm),[51] which substantially decreased the surface defects, enabling a better charge balance in the QLED device (Figure S3, Supporting Information). The synthesized LiZnMgO nanoparticles were also mixed with polyvinylpyrrolidone (PVP) (5 wt%) prior to deposition to improve the morphology, reduce the conductivity, and passivate the quenching sites of the nanoparticles.[52]

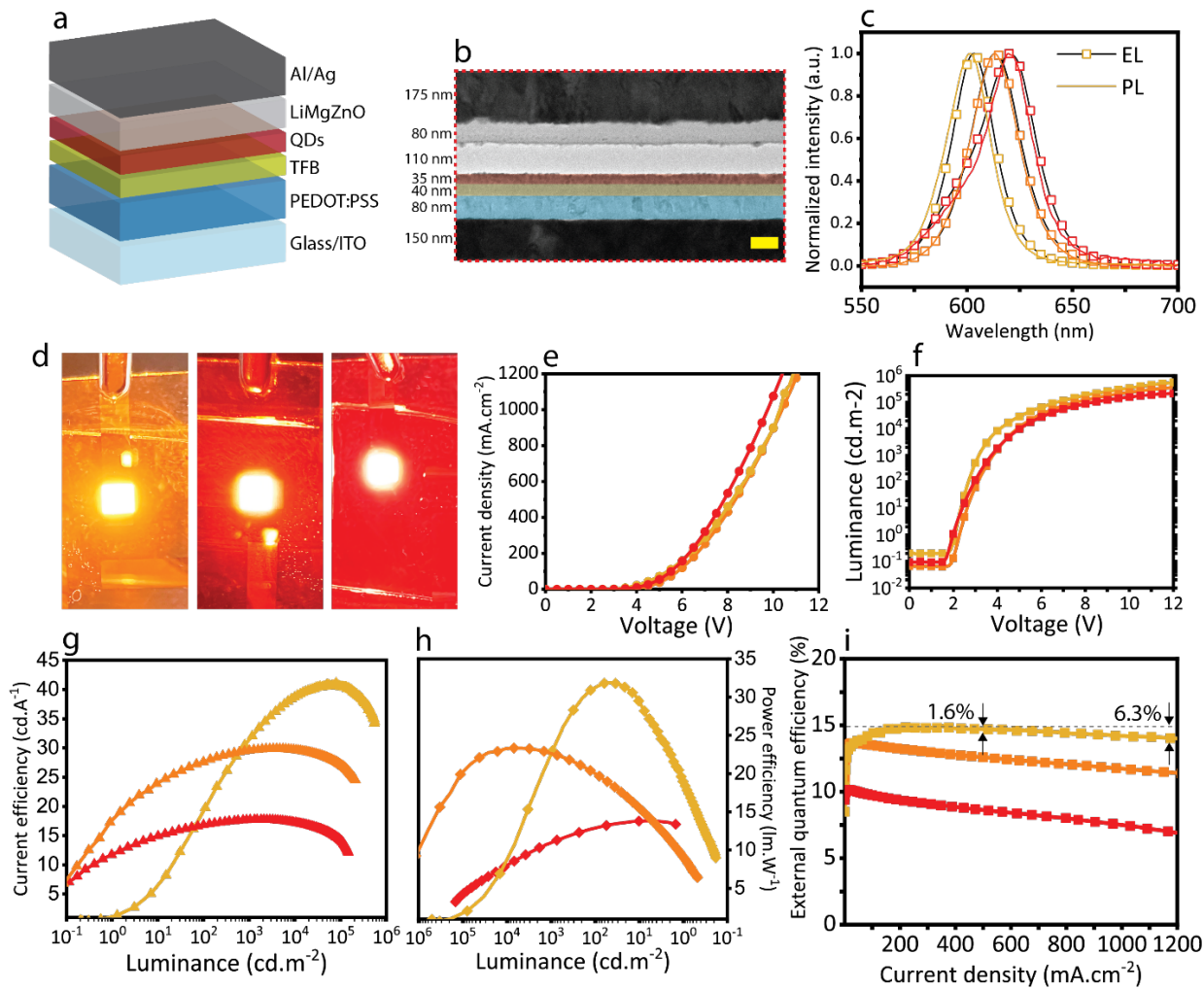

**Figure 4.** (a) Schematic of the QLED devices. (b) Cross-sectional TEM image of a device with experimentally optimized architecture (scale bar = 100 nm). (c) The PL of amber-, orange-, and red-emitting QDs and their EL spectra in different devices. (d) Photographs of the QLED devices with different emission colors. (e) Current density vs. voltage, (f) luminance vs. voltage, (g) current efficiency vs. luminance, (h) power efficiency vs. luminance, and (i) EQE vs. current density curves of the fully in-air fabricated QLED devices from amber-, orange-, and red-emitting ZnCdSe/ZnSe/ZnSeS/ZnS QDs. In (c-i), the line color of the spectra corresponds to the emission color of the QLED device. The level of efficiency droop is calculated based on the best performing device (14.8%) at the current density levels of 500 and 1200 mA.cm$^{-2}$.

To compare the performance of the QLED devices at different wavelengths, three devices were fabricated from the synthesized amber- ($\lambda_{PL}$= 596 nm), orange- ($\lambda_{PL}$= 610 nm), and red-emitting ($\lambda_{PL}$= 620 nm) QD solutions. The PL of the synthesized QDs and the electroluminescence (EL) of the fabricated devices were approximately matched ($\Delta\lambda$ < 3 nm) with a pure color in each wavelength region (Figures 4c and 4d). The (x,y) coordinates of the generated emission in the color tristimulus from amber (A-QLED), orange (O-QLED), and red (R-QLED) devices were (0.6045, 0.3751), (0.6412, 0.3457), and (0.6386, 0.3500), respectively (Figure S5, Supporting

Information). As expected, the 'giant' ZnCdSe/ZnSe/ZnSeS/ZnS QDs with robust luminescent properties enabled excellent performance in QLED devices fully fabricated in air (Figure 4e-i). The turn-on voltages were measured to be 2.3 V, 2.2 V, and 2.0 V for A-QLED, O-QLED, and R-QLED devices, respectively (Figure 4f). The significantly low current density under the turn-on voltage in all the devices indicated a low leakage current due to the optimized charge balance (Figure 4e). Advantageously, the optimized core/multi-shell composition of the QDs and the device architecture enabled achieving an ultra-high luminance level of 560,528 cd.m$^{-2}$ at 12 V for A-QLED, which positions our devices as one of the brightest QLEDs in the amber-to-red wavelength region.[29] The investigation of the current efficiency (CE) and power efficiency (PE) of A-QLED showed maximum values of 41.1 cd.A$^{-1}$ at the luminance level of 79,497 cd.m$^{-2}$ and 31.8 lm.W$^{-1}$ at the luminance level of 2,226 cd.m$^{-2}$, respectively. The highest measured EQE of the A-QLED, O-QLED, and R-QLED devices were 14.8%, 13.7%, and 10.1%, respectively. The EL characteristics of the fabricated devices are summarized in Table 1.

**Table 1.** QLED device parameters of amber, orange and red QLEDs.

| QLED | $\lambda_{PL}$(nm) | $\lambda_{EL}$(nm) | $V_{ON}$ (V) | L at 12 V (cd.m$^{-2}$) | EQE (%) | PE (lm.W$^{-1}$) | CE (cd.A$^{-1}$) |
|---|---|---|---|---|---|---|---|
| A-QLED | 596 | 599 | 2.3 | 560528 | 14.8% | 31.8 | 41.1 |
| O-QLED | 610 | 611 | 2.2 | 357238 | 13.7% | 23.3 | 29.9 |
| R-QLED | 620 | 622 | 2.0 | 216099 | 10.1% | 17.7 | 13.8 |

We believe that the high-performance levels of the fully in-air fabricated QLED devices are due to (i) the high film PLQYs of the synthesized QDs, which increases the internal quantum efficiency (IQE) (IQE = PLQY × charge balance efficiency), (ii) the strong confinement of the generated charge carriers inside the core, which is achieved by lattice-mismatch-engineered growth of different shell compounds, (iii) the ultra-thick shell (14 ML), which increases the aerobic stability of the QDs and the fabricated QLED devices by inhibiting the diffusion of moisture and oxidative agents, and (iv) tailoring the optical and structural properties of the synthesized LiZnMgO nanoparticles to improve the charge balance and effectively passivating the surface defects.

Noticeably, the combination of optimal device structure together with careful QD-level engineering of charge injection in our QLED devices led to the significant suppression of the non-radiative Auger decays, which was evidenced by the negligible EQE droop in a wide range of current densities (Figure 4i). For example, A-QLED device with the highest EQE of 14.8% showed a relative EQE reduction of 1.6% (down to 14.6%) and 6.3% (down to 13.9%) at current densities of 500 and 1200 mA.cm$^{-2}$, respectively. The low EQE droop at high current densities, at which the QDs become more negatively charged, is critical for pushing the barriers towards practical QLED applications in future lighting systems. The QLED devices also showed a T$_{95}$ of 24 h with an initial luminance of 5000 cd.m$^{-2}$, which further highlights their high operational stability (Figure S7, Supporting Information).

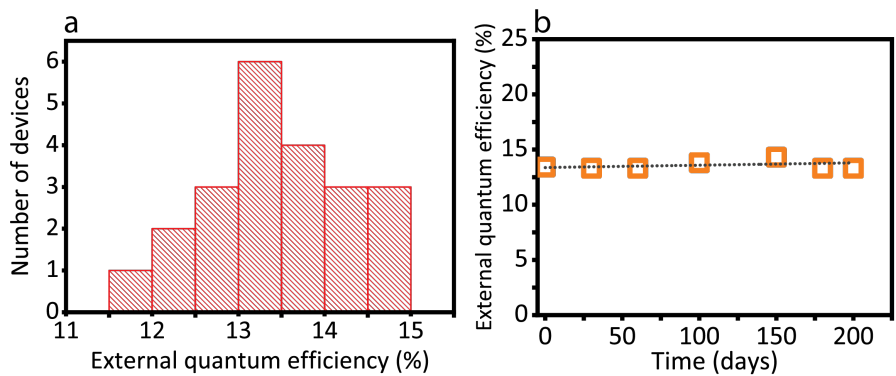

**Figure 5.** (a) Statistical distribution of the EQE values obtained from 22 devices. (b) Measured EQE values of devices fabricated using the same QDs batch stored for 200 days. The linear fitted dashed line shows an EQE$_{avg}$ of 13.7%.

Moreover, the synthesized QDs exhibited remarkable colloidal stability, and when incorporated into the QLEDs after prolonged storage, maintained their EL properties. The ultra-thick shell in the ZnCdSe/ZnSe/ZnSeS/ZnS QD structures significantly lowers the energy transfer between the neighboring QDs and suppresses the non-radiative recombination due to the spatial confinement of the defects from the core. The alloyed composition also reduces the interfacial strains, leading to high PLQY and stability. As a result, it was noted that the performance of the QLED devices fabricated from the same QD batch did not drop significantly over 200 days (Figure 5b). Specifically, the EQE value obtained from amber QLEDs fabricated on day 200 of the QDs solution storage was found to be 13.3%, which was close to the average EQE value of 13.7% obtained from all the amber QLEDs fabricated from the same QD batch.

Finally, enhancing the light outcoupling efficiency of OLEDs and QLEDs (~ 20%) is known to be a subtle approach for further improvement of the device performance. This is particularly important for solid-state lighting applications of these devices where high-brightness levels (at high current densities) are required. Specifically, enhancing the light outcoupling efficiency enables obtaining the same brightness at lower current densities, reducing the Joule heat generated during the device operation. For example, a reliable strategy to improve the light outcoupling efficiency is to reduce the trapped substrate modes by incorporating external high-index optical components on the backside of the device. Therefore, as a simple proof-of-concept approach, to extract more light from our QLEDs, we used a refractive-index-matched immersion oil (n = 1.52) between the QLED substrate and the silicon photodetector employed for the efficiency measurements. As a result, the EQE, luminance, PE, and CE of the same A-QLED device were boosted to 26.6%, 1,028,681 cd.m$^{-2}$, 53.5 lm.W$^{-1}$, and 73.1 cd.A$^{-1}$, respectively, (Figure S6, Supporting Information) comparable to the best values obtained from previously reported state-of-the-art QLEDs fabricated under inert atmosphere. Upon applying the immersion oil, the current densities at the brightness levels of 1000 cd.m$^{-2}$ (2 mA.cm$^{-2}$), 5000 cd.m$^{-2}$ (9.5 mA.cm$^{-2}$), and 10000 cd.m$^{-2}$ (18 mA.cm$^{-2}$) were significantly reduced to 0.31 mA.cm$^{-2}$, 1.18 mA.cm$^{-2}$, and 3.16 mA.cm$^{-2}$, respectively.

## 3. Conclusions

In summary, efficient QLED devices were demonstrated by structural engineering of the synthesized ZnCdSe/ZnSe/ZnSeS/ZnS QDs in the amber-to-red wavelength region. The effective confinement of the charge carriers due to the 14 ML overcoating of lattice-mismatch-engineered ZnSe/ZnSeS/ZnS shells led to the excellent solution and film PLQYs and narrow linewidth of <30 nm within the 595-625 nm spectral region. The formation of an ultra-thick shell inhibited the diffusion of moisture and oxidative species into the core, resulting in the preservation of the optical properties of the QDs for at least 200 days. The superior aerobic stability of the synthesized QDs enabled fully fabricated in-air QLED devices with a maximum EQE of 14.8%, a CE of 41.1 cd.A$^{-1}$, and a luminance of 560,528 cd.m$^{-2}$. Moreover, the significant suppression of Auger recombination led to an extremely low efficiency droop of 1.6% and 6.3% at 500 and 1200 mA.cm$^{-2}$, respectively. Employing a refractive-index-matched immersion oil boosted the EQE and luminance levels of the QLEDs up to 26.6% and >1,000,000 cd.m$^{-2}$, respectively. These results demonstrated the most efficient and brightest QLED devices fabricated under ambient air conditions and could open the door for the realization of efficient QLED devices for future high-brightness lighting systems.

# 4. Experimental Section

**Materials.** Cadmium oxide (CdO, 99.99%, trace metals), trioctylphosphine (TOP, 90% technical grade), sulfur (S, 99.98%), 1-octadecene (ODE, 90%, technical grade), 1-octanethiol (>98.5%), trimethylammonium chloride (TMACl, >98%), potassium hydroxide (KOH, 99.99%), dimethyl sulfoxide (DMSO, >99.9%), magnesium acetate tetrahydrate (99%), zinc acetate dihydrate (>98%), lithium acetate (99.95%), 1-butanol (anhydrous, 99.8%), and polyvinylpyrrolidone (PVP10) with an average molecular weight of $M_w=10000$ were purchased from Sigma-Aldrich. Zinc acetate anhydrous (+99.9%) and ethyl acetate (>99.5%, ACS certified) were purchased from Thermoscientific. Selenium (Se, 99.999%, metals basis) and oleic acid (90%, technical grade) were purchased from Alfa-Aesar. Octane (+ 99%, extra pure) was purchased from Acros Organics. Poly(ethylene dioxythiophene): polystyrene sulfonate (PEDOT: PSS) was purchased from Ossila. Poly[(9,9-dioctylfluorenyl-2,7-diyl)-co-(4,4'-(N-(p-butylphenyl)) diphenylamine)] (TFB) was purchased from American Dye Source. Patterned ITO-glass substrates with a sheet resistance of 15 Ω were purchased from Luminescence Technology Corp. All the reagents were used as received without further purification. The device active area was 4 mm$^2$.

**Synthesis of amber-to-red-emitting QDs.** Preparation of precursor solutions: Se precursor for quick injection: Se was dissolved in TOP under Ar flow. The solution was sonicated to complete the dissolution of powder in TOP and form a transparent colorless solution.

Se precursor for slow injection: Se was dissolved in TOP and ODE under the Ar flow for preparation of the TOPSe solution. The solution was sonicated to complete the dissolution of powder in TOP and form a transparent colorless solution.

S precursor (1): S was dissolved in TOP and ODE under the Ar flow. The solution was heated and sonicated to complete the dissolution of powder in TOP and form a transparent colorless solution.

Se-S precursor: Se precursor (for slow injection) and S precursor (1) were mixed under the Ar flow.

S precursor (2): S precursor (1) was added to ODE and octanethiol solution under Ar flow.

Synthesis of ZnCdSe core QDs: CdO (the amount is defined based on the desired PL wavelength), zinc acetate. oleic acid, and ODE were mixed and degassed. After the evacuation process was finished, the solution was heated. Se precursor (for quick injection) was swiftly injected into the main reaction.

ZnSe shell formation: After the core growth time, TOPSe precursor solution (for slow injection) was injected into the reaction.

ZnSeS shell formation: Se-S precursor was injected into the reaction using a syringe pump.

ZnS shell formation. S precursor (2) was injected into the solution by using a syringe pump.

1-Octanethiol ligand exchange procedure: After finishing the reaction, octanethiol solution was injected dropwise into the reaction. Then, the reaction was heated and cooled down to RT. The purification was repeatedly performed by the addition of reagent alcohol, discarding the supernatant, and re-dissolving the QDs in chloroform. The final step of purification was performed with reagent alcohol and the QD precipitate was re-dissolved in octane for making the devices and characterizations.

**Synthesis of co-doped LiZnMgO NPs.** TMAH precursor was synthesized by dissolving 2.2 g of TMACl in 14 ml of reagent alcohol and 1.1 g of KOH in 16 ml of reagent alcohol, separately. After the complete dissolving, the KOH and TMACl solutions were mixed and centrifuged at 5000 rpm for 3 min. The resulting TMAH was filtered using 0.22 μm PTFE filters, while the KCl solid part was separated. 10%Li:10%Mg:ZnO nanoparticles were synthesized by solution-precipitation method with modifications.[53] In a typical synthesis procedure, 8 mmol of zinc acetate dihydrate was mixed with 1 mmol of magnesium acetate dihydrate and 1 mmol of lithium acetate hydrate. The mixture was dissolved in 25 ml of DMSO and the temperature of the solution was reduced to below 2 °C. At this temperature, 21 ml of TMAH was added dropwise to the reaction. The solution was stirred for 2 h while the temperature was constantly kept below 2 °C. After finishing the reaction, the synthesized nanoparticles were washed twice with ethyl acetate and redispersed in butanol.

**QLED device fabrication.** Indium tin oxide (ITO) patterned glass substrates were cleaned by sonication in a bath of detergent and deionized water, acetone, and iso-propanol, each for 15 min, consecutively. The substrates were then dried with a nitrogen gun, and treated in a UV-ozone system for 15 min. PEDOT:PSS (Heareus HIL E-110) was spin-coated at 5000 rpm for 40 sec and annealed at 130 °C for 20 min. Next, a TFB solution (8 mg.ml$^{-1}$ in toluene) was spin-coated at 3000 rpm and annealed at 110 °C for 20 min. Amber-to-red-emitting ZnCdSe/ZnSe/ZnSeS/ZnS QD solutions (20 mg.ml$^{-1}$ in octane) were spin-coated at 3000 rpm and annealed at 80 °C for 15 min. A solution of 30 mg.ml$^{-1}$ Li:Mg:ZnO and 1.5 mg.ml$^{-1}$ PVP was then spin-coated at 3000 rpm and annealed at 80 °C for 30 min. Finally, a cathode of 50 nm Al (0.25 nm.s$^{-1}$) and 120 nm Ag (0.1-0.2 nm.s$^{-1}$) was deposited by thermal evaporation. Finally, the devices were encapsulated with UV-curable epoxy and a glass slide. The active area of the devices was 4 mm$^2$. During the QLED fabrication procedure, the lab temperature was approximately 22 ± 2 °C (depending on the seasonal conditions) with a humidity level of 11-15% and 25-35% in cold and warm seasons, respectively.

**Instrumentation and characterization.** The absorbance measurements were performed by Agilent Technologies Cary-60 UV-Vis device. For solution and film PLQY measurements, Labspehere and IC2 StellarNet Inc. integrating spheres were used, respectively. The StellarNet spectrometer (CXR-SR-100), optical fiber, and SpectraWiz software were used for data analysis. The standard 1×1 cm$^2$ quartz cuvettes were used for absorbance, PL, and solution PLQY measurements. film PLQY samples were prepared by spin-coating QD solution onto the quartz substrates and annealed for 20 minutes at 80 °C for complete drying. The absorbance of the excitonic peak in solution and film samples in PLQY measurements was set below 0.1. All the PLQY measurements were performed from at least three samples. Time-resolved photoluminescence (TRPL) measurements of core/multi-shell QDs were acquired using excitation from a broadband, pulsed fiber laser source, which was spectrally filtered to provide selective excitation at λ: 436 nm (Fianium WhiteLase, SC4X0 Series). Data was collected using a single photon avalanche diode (Micro Photon Devices, SPD-050-CTD) and PicoQuant PicoHarp 300 event timer. The emission setup used a collimating lens to collect fluorescence, which was coupled through free space to the detector. A dielectric long-pass filter (Omega Optical, RPE500LP) was used to reject laser scatter. All TRPL measurements were acquired at an excitation repetition rate of 5 MHz with an intrinsic time resolution of 4 ps and excitation density of 10 pJ.cm$^{-2}$.pulse$^{-1}$, unless otherwise noted. The XRD samples were prepared by drop casting QD solutions onto a silicon substrate and annealing the substrate at 200 °C for 10-15 mins to ensure the complete evaporation of the solvent. A Bruker D8 diffractometer with a step size of 0.03 and scan speed of 4 sec.step$^{-1}$ was used to perform the XRD measurements at RT. TEM lamellas were prepared on a Thermo-Fisher Helios Hydra Plasma FIB/SEM dual beam system. 30 kV Xe beam was utilized to polish the thin section. TEM and S/TEM/EDX analysis was performed on a JEOL ARM 200cf microscope, which is equipped with a cold-field emitter and probe Cs corrector and operates at 200 kV acceleration voltage.

**Supporting Information**

Supporting Information is available.

**Author Contribution**

S.S. and A.S. conceived the idea and designed the experiments. S.S. and S.D. synthesized the colloidal QDs and optimized their structural and optical properties. S.M.Z. synthesized the LiZnMgO NPs, fabricated the QLED devices, and analyzed their performance metrics. C. J. I. and S. K.-C. performed the time-resolved measurements and analyzed the data. F.T. contributed to the device measurement and analysis. S.S. wrote the first draft of the manuscript. All authors have approved the final version of the manuscript. A.S. and M.P. supervised the project.

**Acknowledgments**

M.P. acknowledges MITACS accelerate program (project no. IT17040) for financial support. S.K.-C. acknowledges support from the Canada Research Chairs Program. The authors sincerely thank Peng Li, Xuehai Tan, and Shihong Xu from NanoFab (University of Alberta) for TEM-EDAX, cross-sectional TEM, and XRD measurements.

**Conflict of Interest**

The authors have filed a patent with PCT application (PCT/CA2022/050550 – April 8, 2022), covering the work presented in this paper.

# Structural Engineering of Colloidal Quantum Dots: Towards Realization of Highly Efficient, Aerobic-Stable, and Droop-Free QLEDs

## Supporting Information

*Sadra Sadeghi, Saeedeh Mokarian Zanjani, Sergey Dayneko, Christian J. Imperiale, Francesco Tintori, Stéphane Kéna-Cohen, Afshin Shahalizad, and Majid Pahlevani\**

S. Sadeghi, S. Mokarian Zanjani, F. Tintori, and M. Pahlevani
Department of Electrical and Computer Engineering
Queen's University
Kingston, Ontario K7L 3N6, Canada
E-mail: majid.pahlevani@queensu.ca

A. Shahalizad and S. Dayneko
Printed Electronics Division
Genoptic LED Inc.
Calgary, Alberta T2C 5C3, Canada

C. J. Imperiale and S. Kéna-Cohen
Department of Engineering Physics
École Polytechnique de Montréal
PO Box 6079, succ. Centre-Ville, Montreal QC H3C 3A7, Canada

S. Sadeghi and S. Mokarian Zanjani contributed equally to this work.

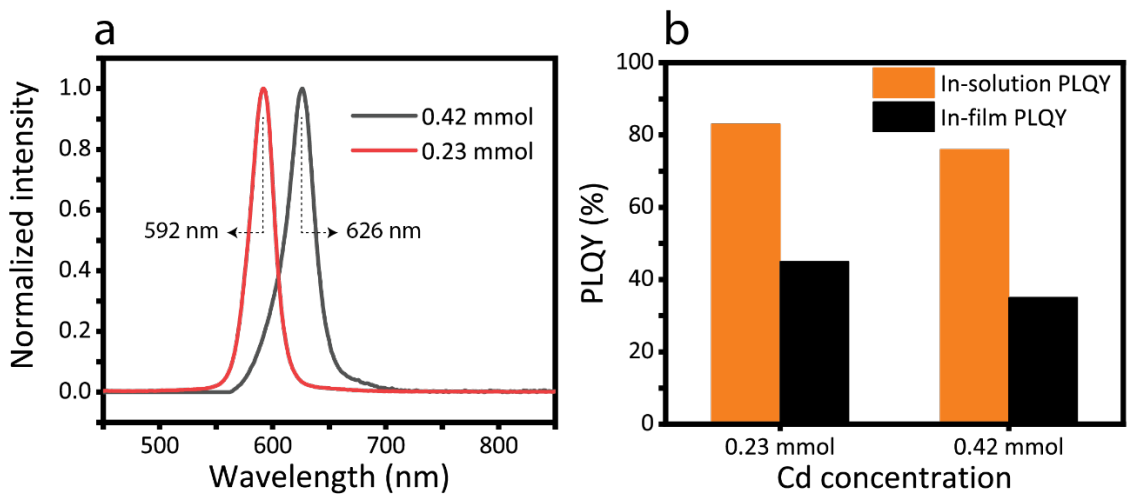

**Figure S1.** (a) The photoluminescence spectra, (b) solution, and film PLQYs of the synthesized ZnCdSe/ZnSe/ZnSeS/ZnS QDs with different concentrations of the Cd precursor.

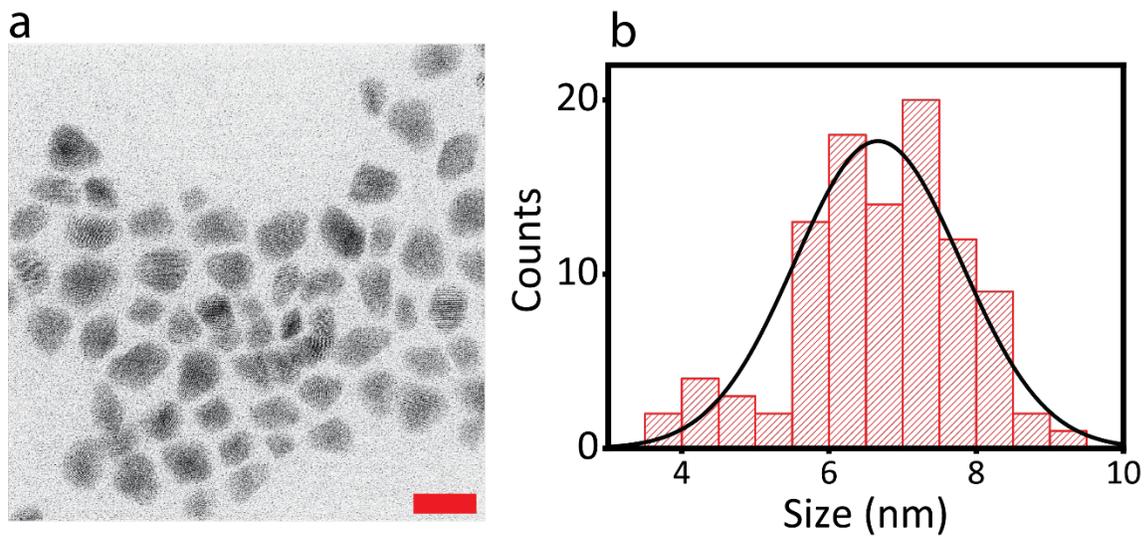

**Figure S2.** (a) TEM image of the ZnCdSe core QDs (scale bar = 10 nm). (b) Size distribution of the synthesized core QDs taken out of 100 particles. The average size of the synthesized core QDs was 6.7 nm ± 1.1 nm.

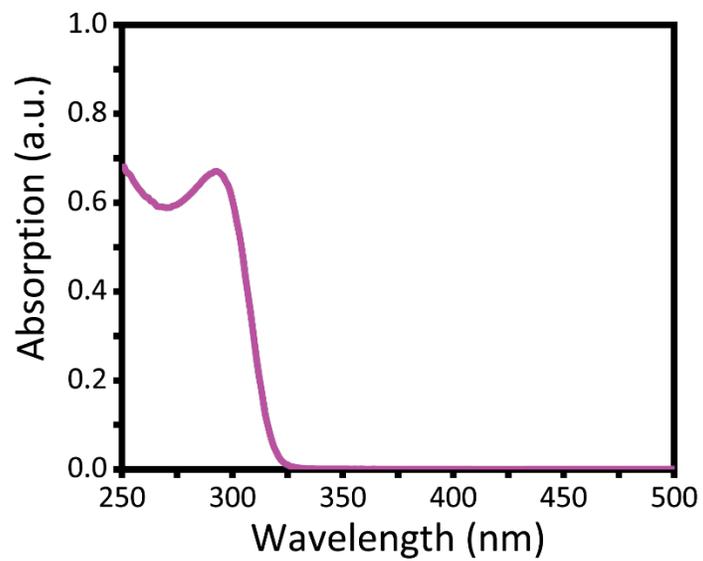

**Figure S3.** The absorbance spectrum of the synthesized LiZnMgO nanoparticles.

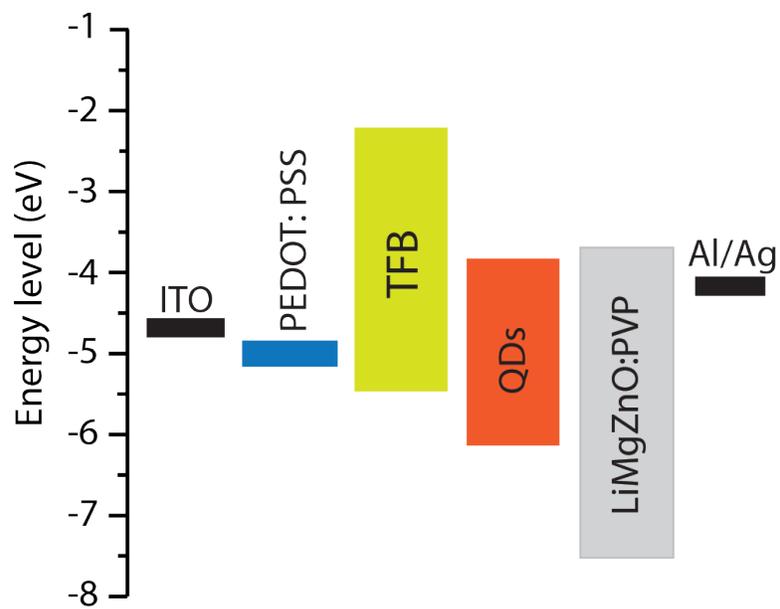

**Figure S4.** The schematic energy levels of materials in the multi-layered QLED device.

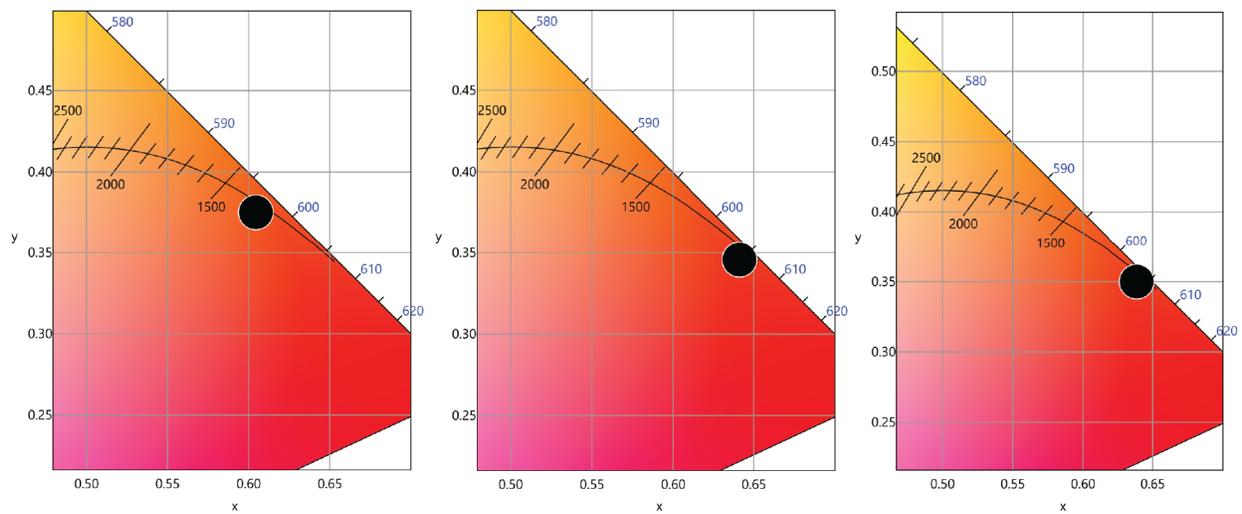

**Figure S5.** The color coordinates of the (left) A-QLED, (middle) O-QLED, and (right) R-QLED devices in tristimulus color coordinates.

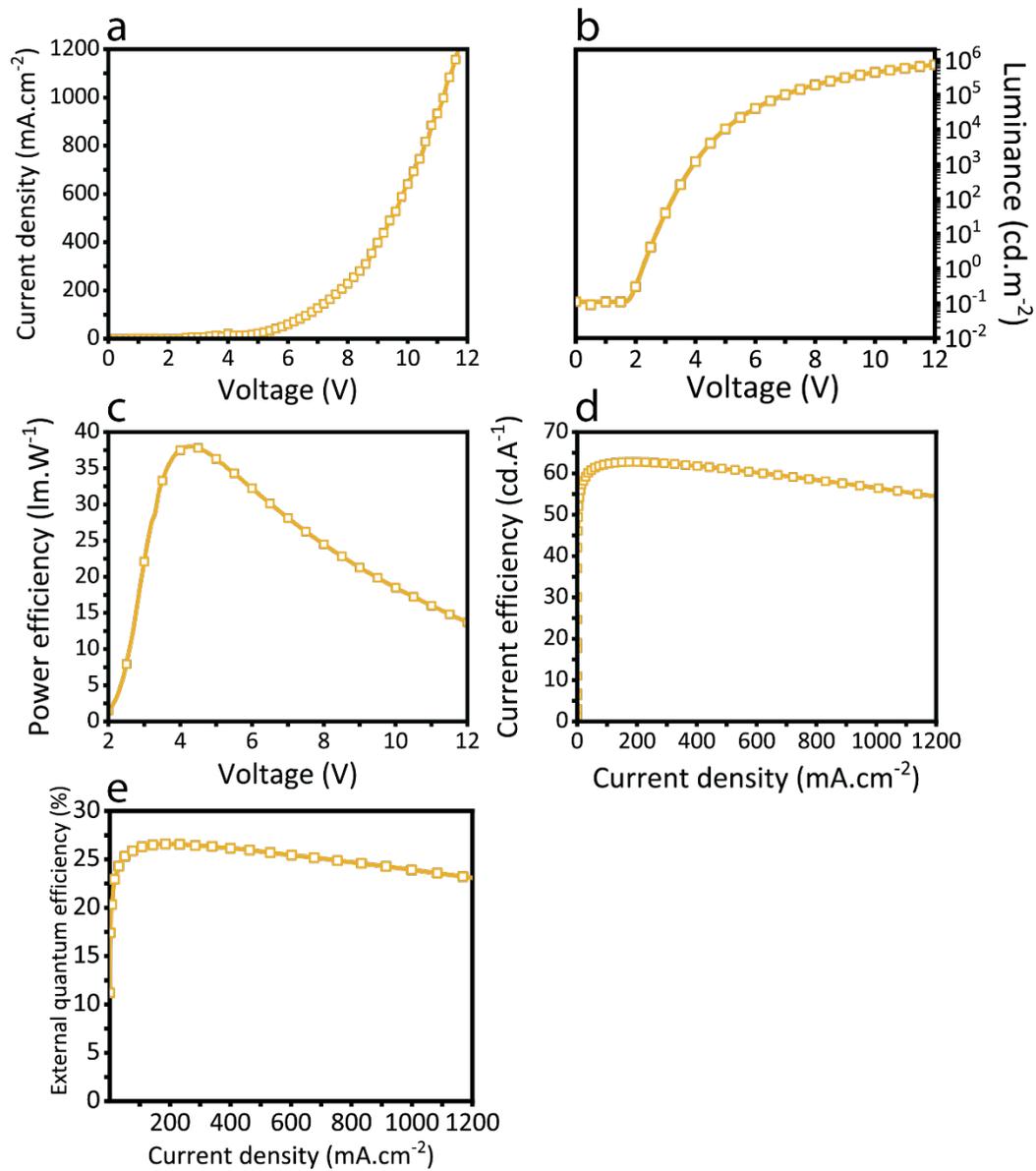

**Figure S6.** (a) the current density vs. voltage, (b) the luminance vs. voltage, (c) the power efficiency vs. voltage, (d) the current efficiency vs. current density, and (e) the EQE vs. current density spectra of the amber-emitting QLED with the refractive-index-matched immersion oil.

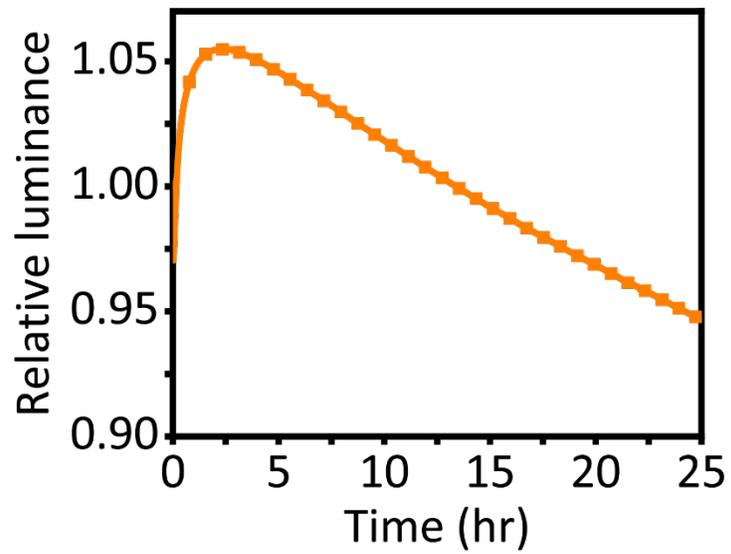

**Figure S7.** The stability measurement of the QLED devices with initial luminescence of 5000 cd.m$^{-2}$.